\documentclass[dvips]{article}

\usepackage{graphics}
\begin{document}
\bibliographystyle{unsrt}
\newcommand {\gtsim } {\,\vcenter{\hbox{$\buildrel\textstyle>\over\sim$}}\,}
\newcommand {\lesssim } {\,\vcenter{\hbox{$\buildrel\textstyle<\over\sim$}}\,}
\newcommand {\BE} {\begin{equation}}
\newcommand {\EE} {\end{equation}}
\newcommand  {\sign }  {\,{\rm sign} \,}
\def\x{{\bf x}}
\def\y{{\bf y}}
\def\p{{\bf p}}
\def\F{{\bf F}}
\def\H{{\cal H}}
\def\A{{\cal A}}
\def\sgn{{\rm sgn}}
\def\abs{{\rm abs}}

\title{Simulated Annealing using Hybrid Monte Carlo}

\author{R. Salazar$^{1}$, R. Toral$^{1,2}$\\
(1) Departament de F\'{\i}sica, Universitat de les Illes Balears\\ 
(2) Instituto Mediterr\'aneo de Estudios Avanzados (IMEDEA, UIB-CSIC)\\
07071 Palma de Mallorca, Spain.}

\date{\today}

\maketitle
\begin{abstract}

We propose a variant of the Simulated Annealing method for 
optimization in the multivariate analysis of differentiable functions.
The method uses global actualizations via the Hybrid Monte Carlo 
algorithm in their generalized version for the proposal of new 
configurations. 
We show how this choice can improve upon the performance of simulated
annealing methods (mainly when the number of variables is large) 
by allowing a more effective searching scheme and
a faster annealing schedule. 
Hybrid Monte Carlo. Multivariate minimization.
 
\end{abstract}

\newpage
\section{Introduction}
\label{intro}

An important class of problems can be formulated
as the search of the absolute minimum of a function of a
{\sl large}  number of variables. These problems include 
applications in different fields such as
Physics, Chemistry,
Biology, Economy, Computer Design, Image processing, etc.\cite{laa88}. 
Although in some occasions, such as the NP-complete class of
problems\cite{gar79},
it is known that no algorithm can surely find the absolute minimum in a
polynomial time with the number of variables, some
very successful heuristic algorithms have been developed. 
Amongst those, the Simulated Annealing (SA) method of  Kirkpatrick, 
Gelatt and Vecchi\cite{kir83}, has proven to be very successful in 
a broad class of situations.
The problem can be precisely defined as finding
the value of the $N$--dimensional vector $\x \equiv (x_1,x_2,...,x_N)$, 
which is an absolute minimum of the real function $E(\x)$. 
For large $N$, a direct search method is not effective due to the large
configuration space available. Moreover, more sophisticated methods, 
such as downhill simplex or those using the gradient 
of $E(\x)$\cite{pre94},
are likely to get stuck in
local minima and, hence, might not able to reach the absolute minimum.

SA is one of the most effective methods devised to overcome these difficulties.
It allows
escaping from local minima through tunnelling and also by accepting higher 
values of $E(\x)$ with a carefully chosen probability\cite{kir83}.
The method is based on an analogy with Statistical Physics:
the set of variables $(x_1,\dots,x_N)$ form the phase space of 
a fictitious physical system.
The function $E(\x)$ is considered to be
the system's energy and the problem is reduced to that of
finding the ground state configuration of the system. 
It is known that if a system is heated to a very high temperature $T$
and then it is {\sl slowly} cooled down to the absolute zero
(a process known as annealing), the 
system will find itself in the ground state. The cooling rate
must be slow enough in order to avoid getting trapped in some metastable
state. At temperature $T$, the probability of being on a 
state with energy $E(\x)$ is given by the Gibbs factor:
\begin{equation}
\label{eq:gibbs}
P(\x) \propto \exp(-E(\x)/T).
\end{equation}
From this relation we can see that high energy states
can appear with a finite probability at high $T$.
If the temperature is lowered,
those high energy states become less probable and, as $T \to 0$, only 
the states near the minimum of $E(\x)$ have a
non--vanishing probability to appear. In this way, 
by appropriately decreasing the temperature
we can arrive, when $T \rightarrow 0$, to the (absolute) minimum energy state.
In practice, the method proceeds as follows: at each annealing step $k$ there
is a well defined temperature $T(k)$ and the system is let to evolve
long enough such that it thermalizes at temperature $T(k)$. The temperature
is then lowered according to a given annealing schedule $T(k)$ and the
process is repeated until the temperature reaches $T=0$.

To completely specify the SA method, one should give a way of
generating representative configurations at temperature $T$, and also
the variation of the temperature with annealing step, $T(k)$.
For the generation of the configurations, the Monte Carlo 
method (MC) is widely used\cite{kal86,her86,bin92}.
MC introduces an stochastic dynamics in the system by proposing
configuration changes $\x \to \x'$
with probability density function (pdf) $g(\x'|\x)$, i.e, if the system
variables adopt presently the value $\x$, the probability that the 
new proposed value lies in the interval $(\x',\x'+d\x')$ is $g(\x'|\x)d\x'$.
This proposal is accepted with a probability $h(\x'|\x)$. 
Much freedom is allowed in the choice of the proposal and acceptance
probabilities.  A sufficient condition 
in order to guarantee that the Gibbs distribution is properly sampled,
is the detailed balance condition:
\begin{equation}
\label{eq:balance}
g(\x' | \x) h(\x' | \x) \exp(-E(\x)/T) = 
g(\x | \x') h(\x | \x') \exp(-E(\x')/T).
\end{equation}
Once the proposal pdf $g(\x'|\x)$ has been conveniently specified, 
the acceptance probability $h(\x'|\x)$ is given as a convenient solution 
of the previous detailed balance equation. Usually (see next section)   
the proposal probability $g(\x'|\x)= g(\Delta \x)$ is a
symmetric function of the difference $\Delta \x \equiv \x'-\x$, $g(\Delta
\x)=g(-\Delta \x)$ and a commonly used
solution to the detailed balance equation is the Metropolis choice:
\begin{equation}
\label{eq:acc}
h(\x' | \x)= \min \left( 1, \exp\left[-\left(E(\x')- E(\x)\right)/T\right]
\right) ,
\end{equation}
although other solutions have been also widely used in the literature.

The various SA methods differ essentially in the choice of the proposal
probability $g(\Delta \x)$ and the annealing schedule $T(k)$. 
One can reason that the cooling schedule $T(k)$ might not be independent of the
proposal probability $g(\Delta \x)$, i.e. $T(k)$ should be chosen 
consistently with the
selected $g(\Delta \x)$ in such a way that the configuration space
is efficiently sampled. In the next section we briefly review the main choices
used in the literature. We mention here that most of them involve only the
change of {\sl one} single variable $x_i$ at a time, i.e. they consist 
generally of
small {\sl local} moves. $N$ of these local moves constitute what is called a 
Monte Carlo Step (MCS). The reason for using only local moves is that 
the acceptance probability given by (\ref{eq:acc}) is very small if all
the variables are randomly changed at once, because the change in energy
$E(\x')-E(\x)$ is an extensive quantity that scales as the number of variables
$N$. Hence, the acceptance probability near a minimum of $E(\x)$ becomes
exponentially small. Since $\Delta \x$ is a small quantity, the cooling
schedule must be consequently small, because a large cooling rate would not
allow the variables to thermalize at the given temperature.
It is then conceivable that the use of a global update scheme could 
improve upon the existing methods by allowing the use of larger cooling
rates.

In this paper we investigate the effect of such a global update dynamics. 
Specifically, we use the Hybrid Monte Carlo (HMC) algorithm\cite{dua87}
for the 
generation of the representative configurations at a given temperature. 
By studying some examples, we show that the use
of this global dynamics allows quite generally 
an exponentially decreasing cooling schedule, which is the best one can 
probably reach with other methods. Another advantage of the use of the HMC
is that the number of evaluations of the energy function $E(\x)$ is greatly
reduced. Finally, we mention that  
the use of a generalized HMC \cite{tor94,fer93} allows to treat efficiently
minimization problems in which the range of variation is different for 
each variable.

The rest of the paper is organized as follows: in section II we 
briefly review some of the existing SA methods; in section III we 
explain how to implement Hybrid Monte Carlo in an optimization problem;
in section IV we use some standard test functions to compare our method
with previous ones; and
in section V we end with some conclusions and outlooks.

\section{Review of Simulated Annealing Methods}

Amongst the many choices proposed in the literature,
we mention the following:

\noindent -{\it Boltzmann Simulated Annealing} (BSA)\cite{ger84}:
Based on a functional form derived for many physical systems belonging to the
class of Gaussian-Markovian systems, at each annealing step $k$ 
the algorithm chooses a proposal
probability given by local moves governed by a Gaussian distribution:
\begin{equation}
g(\Delta \x) \sim \exp\left[-\frac {|\Delta \x|^2} {2 T(k)}\right] .
\end{equation}
The Metropolis choice 
(\ref{eq:acc}) is then used for the acceptance. This choice for the
proposal probability and the use of purely local moves
imply that the annealing schedule must be particularly slow: 
$T(k)=T_0/ \ln(1+\lambda k)$, for some value of the cooling rate $\lambda$.

\noindent -{\it Fast Simulated Annealing} (FSA)\cite{szu87}:
States are generated with a proposal probability that has a Gaussian--like
peak and Lorentzian long--range tails that imply occasional 
long jumps in configuration space.
These eventual long jumps make
FSA more efficient than any algorithm based on any bounded 
variance distribution (in particular, BSA).
The proposal probability at annealing step $k$ 
is a $N$--dimensional Lorentzian distribution:
\begin{equation}
g(\Delta \x) \sim T(k)(|\Delta \x|^2+T(k)^2)^{-\frac{N+1}{2}}.
\end{equation}
One of the most significant consequences of this choice is that it
is possible to use a cooling schedule 
inversely proportional to the annealing step $k$, 
$T(k)=T_0/(1+\lambda k)$, which is exponentially faster than the BSA.

\noindent -{\it Very Fast Simulated Reannealing} (VFSR)\cite{les92}:
In the basic form of this method, the change $\Delta \x$ is generated 
using the set of random variables $\y\equiv (y_1,\dots,y_N)$
\begin{equation}
\Delta x_i=(B_i - A_i) y_i, 
\end{equation}
($A_i$ and $B_i$ are the minimum and maximum value of the $i$--th dimension
range). The proposal probability is defined as
\begin{equation}
g(\y) = \prod_{i=1}^N \frac 1 {2( \mid y_i \mid + T_i(k)) 
\ln(1+1/T_i(k))}.
\end{equation}
Notice that different temperatures $T_i(k)$ can be in principle
used for the updating of different variables $x_i$.
For the acceptance probability, one uses the Metropolis
choice (\ref{eq:acc}) with yet another temperature $T_0(k)$.
This proposal allows the following annealing schedule: 
$T_i(k)=T_i(0) \exp(-\lambda_i k^\frac 1 N)$, $i=0,1,\dots,N$, 
which is not very efficient for large
number of variables $N$. 
A more detailed description of the VFSR algorithm can be found in
\cite{les92}.

\noindent -{\it Downhill Simplex with Annealing} (DSA)\cite{pre94}:
This method combines the Downhill Simplex (DS)
method (which is basically a searcher for local minima) 
with a Metropolis like
procedure for the acceptance. The DS samples the configuration
space by proposing moves of the ``simplex". A simplex being
a geometrical figure with $N+1$ vertices 
in the $N$--dimensional phase space. The moves are usually
reflections, expansions, and contractions. The acceptance
part is implemented by adding logarithmically distributed
random variables proportional to the temperature to the energy
before the move and subtracting a similar random variable after the
move. The move is accepted if the energy difference is negative.
According to reference 
\cite{pre94} different annealing  schedules $T(k)$ should 
be used for different problems. In the implementation we have made of this
method (see section IV) an exponential decay has been used.

\section{Hybrid Simulated Annealing}

The alternative method we propose --Hybrid Simulated Annealing (HSA)--
uses the Hybrid Monte Carlo (HMC)\cite{dua87} in their generalized
version\cite{tor94,fer93} to generate the representative
configurations. We first review the HMC method.

In its simplest and original form, HMC introduces
a set of auxiliary momenta variables ${\p}\equiv (p_1,\dots,p_N)$ and
the related Hamiltonian function $\H(\x,\p)$:
\begin{equation}
\label {eq:ham}
\H(\x,\p)=E(x_1,\dots,x_N)+ \frac 1 2 \sum_{i=1}^{N} p_i^2=E(\x) + \p^2/2 .
\end{equation}
From the Gibbs factor:
\begin{equation}
\label{eq:mom}
P(\x,\p) \propto \exp[-\H(\x,\p)/T] = \exp[-E(\x)/T] \exp[-\p^2/2T],
\end{equation}
we deduce that, from the statistical point of view, the momenta $\p$ are
nothing but a set of independent, Gaussian distributed, random variables
of zero mean and variance equal to the system temperature $T$.
There is no simple closed form for the proposal 
probability $g(\x'|\x)$,  and the proposal
change $\x \to \x'$ is done in the 
following way: first, a set of initial values 
for the momenta $\p$ are generated by using the Gaussian distribution 
$\exp[-\p^2/2T]$ as suggested by the equation (\ref{eq:mom}); next, Hamilton's 
equations of motion, $\dot x_i=p_i$, $\dot p_i = F_i$, 
where $F_i(\x)=-\partial E(\x)/\partial x_i$ is the ``force" acting on
the variable $x_i$, 
are integrated numerically using the {\sl leap--frog}
algorithm with a time step $\delta t$:
\begin{eqnarray}
\label{eq:leap1}
x_i' & = & x_i + \delta t p_i + \frac {\delta t^2}{2} F_i(\x) \\ \nonumber
p_i' & = & p_i + \frac {\delta t}{2} [ F_i(\x) + F_i(\x')],\hspace{1.5cm}
i=1,\dots,N.
\end{eqnarray}
The proposal $\x'$ is obtained after $n$ iterations of the previous 
basic integration step. In other words: by numerical integration of Hamilton's
equations during a ``time" $n \delta t$.
The value $\x'$ must now be accepted with a probability given by:
\begin{equation}\label{eq:acc2}
h(\x'|\x)=\min\left(1,\exp\left[-\left(\H(\x',\p')- 
\H(\x,\p)\right)/T\right]\right).
\end{equation}
Notice that this acceptance probability uses the total Hamiltonian function
$\H(\x,\p)$ instead of simply the function $E(\x)$ as in the 
methods of last section 
(compare (\ref{eq:acc2}) and (\ref{eq:acc}))\cite{com2}. 
Although Hamilton's equations exactly conserve the energy $\dot \H=0$,
the difference $\Delta \H \equiv \H(\x',\p')- \H(\x,\p)$
is not equal to zero due to the finite time step discretization errors
and one has quite generally $\Delta \H=O(N\delta t^l)$ for
some value of $l$. In this way, although the mapping is a global one, i.e.
all the variables are updated at once, it is still possible to have
an acceptance probability of order unity by properly choosing the time step 
$\delta t$ and one can have large changes in phase space
at a small cost in the Hamiltonian. 
Notice that the Hamiltonian
difference $\Delta \H$ being small, does not necessarily imply that 
$\Delta E$ is small and once can in principle accept moves which imply a large
change in the energy $E(\x)$. 

In order to generate configurations at temperature $T$, one still must
satisfy the detailed balance condition, equation (\ref{eq:balance}).
One can prove that 
sufficient requirements for this detailed balance condition
to hold are that the mapping given by eqs.(\ref{eq:leap1}) satisfies time 
reversibility and area preserving\cite{com3}. These
two properties are exactly satisfied by Hamilton's equations and are also
kept by the leap--frog integration scheme.
Under those conditions, the Gibbs 
distribution (\ref{eq:gibbs}) for the original variables $\x$
is properly sampled. 
It is possible to further generalize the HMC method by 
using more general mappings
satisfying the conditions of time reversibility and area preserving. 
In reference \cite{fer93} it was shown that those conditions were 
satisfied by the mapping induced by $n$ iterations of the
following basic step:
\begin{eqnarray}
\label{eq:leap2}
x_i' & = & x_i + \delta t \sum_{j=1}^{N} A_{ij} p_j + \frac {\delta t^2}{2} 
\sum_{j,k=1}^{N} A_{ik} A_{jk} F_j(\x), \\ \nonumber
p_i' & = & p_i + \frac {\delta t}{2} \sum_{j=1}^{N} A_{ji}[ F_j(\x) + F_j(\x')],
\hspace{2.0cm} i=1,\dots,N,
\end{eqnarray}
where $A_{ij}$ is an arbitrary matrix. This mapping can be thought as
the leap--frog numerical integration of the following equations of motion:
\begin{eqnarray}
\dot x_i & = & \sum_{j} A_{ij}p_j, \\ \nonumber
\dot p_i & = & \sum_{j} A_{ji} F_j.
\end{eqnarray}
An straightforward calculation shows that these equations, although not 
being Hamiltonian, still conserve energy, $\dot \H =0$, and the main
features mentioned above of the standard HMC method are still maintained.
Convenient choices for matrix $A_{ij}$ are: diagonal in Fourier space (Fourier
acceleration), or a diagonal
matrix: $A_{ij} =A_i \delta_{ij}$. This last choice
allows an effective integration time
step $\delta t_i = \A_i \delta t$ different for each variable (compare
with (\ref{eq:leap1})):
\begin{eqnarray}
\label{eq:leap3}
x_i' & = & x_i + \delta t_i p_i + \frac {\delta t_i^2}{2} F_i(\x), \\ \nonumber
p_i' & = & p_i + \frac {\delta t_i}{2} [ F_i(\x) + F_i(\x')],\hspace{1.0cm}
i=1,\dots,N
\end{eqnarray}
The possibility of using different time steps for each variable accounts
for the fact that the range of variation might differ for each variable.
This is the case, for instance, of Corana's function (see next section).

Summing up, the HMC proceeds by generating representative configurations
by using a proposal obtained by some of the mappings given above. 
This proposal must now be accepted with a probability given by (\ref{eq:acc2}).
In this paper, we have used mainly the basic mapping given by (\ref{eq:leap1})
except in one case (Corana's function) in which the mapping (\ref{eq:leap3})
has been used instead.
The temperature must then be decreased towards zero as in other SA methods.
Notice that in the case $T = 0$ the random component of the evolution
(the momenta variables) in Eq.(\ref{eq:leap1}) is zero and then the proposal
coincides with that of gradient methods.

The HMC has been extensively used in problems of Statistical
Physics\cite{tor95}. For our purpose here, we have found that
the use of the previous Hamiltonian based
global update  of the statistical system associated with
the energy $E(\x)$, allows 
a much more effective annealing schedule and
searching scheme than,
for instance, the Boltzmann, Fast annealing and Very Fast Reannealing
methods mentioned above. In
particular we have been able to use quite generally an exponential annealing
schedule: $T(k)=T_0 e^{-\lambda k}$.
Moreover, since in HMC the acceptance decision is taken after 
all the $N$ variables have been updated, the number of energy function
evaluations is greatly reduced. This turns out to be important in those
problems in which the calculation of the energy function $E(\x)$ takes
comparatively a large amount of computer time.

\section{Results}
In order to compare our algorithm with the different ones
proposed in the literature, we have
used a set of five test functions: a multidimensional
paraboloid, a function from De Jong's test\cite{jon81}, Corana's
highly multi--modal function\cite{cor87} and two other functions
with many local minima. We now define and describe in some detail 
these functions.

The first function, $f_1(\x)$,
is a $N$--dimensional paraboloid:
\begin{equation}
f_1(\x)=\sum_{i=1}^N x_i^2.
\end{equation}
Here we use the test value $N=200$ and
to compare with the results in  \cite{dyk94}, we also use the value $N=3$. 
Although this is a particularly simple function with a single minimum
$f_1=0$ located at $x_i=0$, $i=1 \dots N$, it 
ultimately describes the late stages of the behaviour of the SA 
algorithm when we are near a local or global minimum of any 
differentiable function.

The second function, $f_2(\x)$, is a two dimensional $(N=2)$ function taken
from De Jong's test typically used
for benchmarking Genetic Algorithms\cite{jon81}:
\begin{equation}
f_2(\x)=\left[0.002+\sum_{j=1}^{25} [j + (x_1-a_{j})^6
+ (x_2-b_{j})^6]^{-1}\right]^{-1},
\end{equation}
where the vectors $a$, $b$ have the following 25 components:
\begin{eqnarray}
a_{j} &=& \{-32,-16,0,16,32,-32,-16,0,16,32,\dots,-32,-16,0,16,32\}, 
\nonumber \\
b_{j} &=& \{-32,-32,-32,-32,-32,-16,-16,-16,-16,-16,\dots,32,32,32,32,32\}, 
\nonumber
\end{eqnarray}
this function has $25$ local minima, and the global minimum is
$f_2=0.998004$, at $x_1=x_2=-32$.

The $f_3(\x)$ function is the Corana's function:
\begin{equation}
f_3(\x)=\sum_{i=1}^{N} \left\{
\begin{array}{ll}
0.15~(0.05~\sgn(z_i)+z_i)^2 ~d_i  ~~~~ & {\rm if~} |x_i-z_i| < 0.05 \\
d_i x_i^2 ~~~~&  {\rm otherwise}
\end{array}
\right.
\end{equation}
\begin{equation}
z_i = 0.2~\lfloor | 5~ x_i | +0.49999 \rfloor \sgn(x_i) \nonumber  \\
\nonumber
\end{equation} 
$d_i$ is an $N$--dimensional vector. In our tests (and following
\cite{dyk94} we have used $N=10$ and $d=(1,1000,10,100,1,1000,10,100,1,1000)$.
This function, which 
has many local minima and is discontinuous and piecewise differentiable,
turns out to be one of the most difficult test functions, 
because the different variables have different scales of 
variation. The global minimum is $f_3(\x)=0$, at $x_i=0$, $i=1 \dots N$.

\begin{figure}[htbp]
\hspace{-2.0cm}
\includegraphics{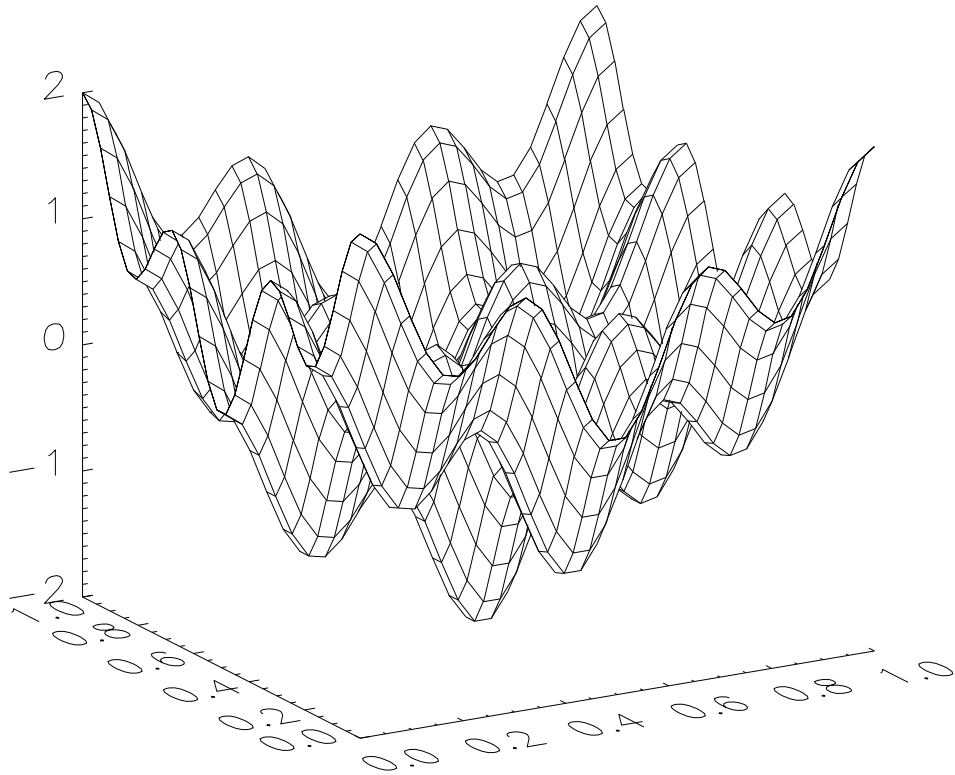}
\caption{\label{fig1} 
Plot of $f_4$ function, equation (\ref{f4}) for $N=2$, in one period. 
Notice the presence of many relative minima, but only one absolute minimum
at $x_1=x_2=0.5$.}
\end{figure}

The $f_4(\x)$ function is defined by:
\begin{equation}
\label{f4}
f_4(\x)=\frac {1}{2N} \sum_{i=1}^N \frac {\sin(4 \pi K x_i)}
{\sin(2 \pi x_i)},
\end{equation}
with $N=200$, $K=2$. 
This function is periodic and has $(2K-1)^N$ local minima per period. The
absolute minima are at $x_i=(2m+1)/2$, $m \in \bf{Z}$, $i=1 \dots N$, and
the minimum value is $f_4(\x)=-K$ (see figure \ref{fig1}).

And, finally, the $f_5(\x)$ function is defined by:
\begin{equation}
f_5(\x)=\sum_{i=1}^N |x_i|^{\alpha} - \prod_{i=1}^N \cos(4 \pi x_i),
\end{equation}
with $N=10$ and $\alpha = 1.3$. Again, this function has many local minima.
The absolute minimum is $f_5=-1$ at $x_i=0$, $i=1 \dots N$.

We present results of the optimization of these typical test functions
performed with the methods described above: Fast Simulated Annealing (FSA), 
Very Fast Simulated Reannealing (VFSR), Downhill Simplex with annealing (DSA)
and the Hybrid Simulated Annealing (HSA).
Amongst other quantities, 
we have focused, as usual in this field, on the number of evaluations
of the function and the CPU time needed to achieve a given accuracy $\epsilon$ 
in the minimum value of each function. These minimum values 
being exactly known for the test functions used.
The results are summarized in tables (\ref{tab1}) and (\ref{tab2}) 
after averaging over $10$ realizations. An accuracy value of $\epsilon=10^{-3}$
has been used, although similar results hold for other values of $\epsilon$.
We have programmed the algorithms for the FSA, DSA and HSA methods,
whereas the results for VFSR have been taken directly from \cite{dyk94}.
For a given test function, we have used 
the same initial condition, $\x_{initial}$, for each method. 
As a general trend, we can see that HSA performs better than the other methods
when the number of variables $N$ is large. This does not imply that HSA 
performs extremely worse for small values of $N$. An important advantage
of HSA in front of other methods is that the number of function evaluations 
is much smaller (in table  (\ref{tab1})
the number of function evaluations includes also the
calculation of the forces necessary in the HSA method).
This might turn out to be very important in those problems
in which the function evaluation takes a long computer time.
We now report in some detail the results of each test function:

As mentioned before, the $f_1$ function, a parabolic function with a single
minimum, serves to model the behavior close to a minimum of 
any function, i.e. the situation for low enough temperature. 
When the number of variables is small, $N=3$, it turns out that the fastest
method (in the sense that it reaches the minimum in less computer time) 
is DSA although HSA needs less function evaluations.
However, when the
number of variables is large, $N=200$, the cost in CPU time and
number of function evaluations is very favorable to HSA.
In general, the performance of the DSA method worsens when the problem
has many minima.
This is obvious when looking at the results for the De Jong's $f_2$,
the Corana $f_3$ and the $f_4$ functions
for which the DSA could not even find the absolute minimum.

The $f_2$ function is another example in which the HSA can not offer 
a better alternative than other methods, stressing the fact again that
for small number of variables the use of a global actualization turns 
out to be irrelevant. In this case, VFSR needs less number of function
evaluations than any other method. However, for large number of
variables $N$, the cooling schedule required for VFSR is necessarily
slow (see the discussion in section 2) making it inefficient for
large $N$.

The functions in which the variables have a wide range of variation 
(for instance Corana's function $f_3$) can
be better handled using the general version of HSA.
Remember that the rescaling in (\ref{eq:leap3})
allows an effective integration time step $\delta t_i = \A_i \delta t$ 
different for each variable. So, one can tune $A_i$ 
to solve efficiently this kind of problems.
In our case, the range of variation of the variables come 
essentially from
the part $V({\bf x})=d_i x_i^2$ of the Corana's function. Then,
from the equations
of motion $\dot x_i=A_i p_i$ and $\dot p_i =A_i F_i$ we have $\ddot x_i=A_i^2
F_i$. The force is $F_i=-2~d_i x_i$ and we have $\ddot x_i=-A_i^2 d_i x_i $ ,
so we chose $A_i= \frac 1 {\sqrt d_i}$ in order that each variable
has the same effective time scale for evolution.

\begin{figure}[htbp]
\hspace{-2.0cm}
\includegraphics{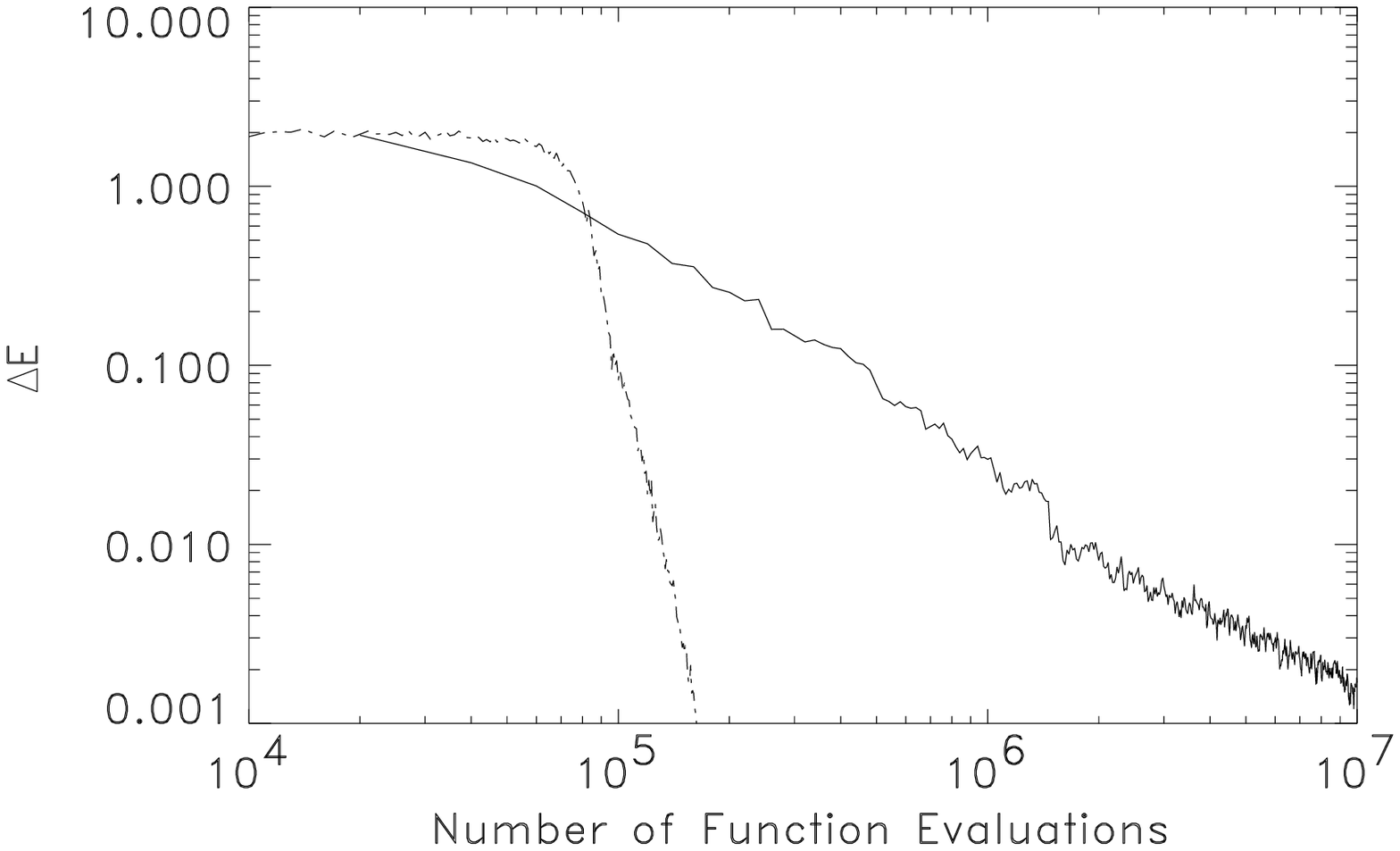}
\caption{\label{fig2}
Plot of "energy" difference with respect to the ground state value, 
versus the number of function evaluations, for the $f_4$ function (\ref{f4})
with $N=200$ using HSA (dotted line) and FSA (continuous line),
both initialized in $\x_{initial}=1.0$, 
the other parameters have the following values,
for FSA: $T_0=0.8$, $m=100$, $\lambda=100$; and for HSA: $T_0=1.0$, $m=10$,
$n=10$, $\delta t=0.3$ $\lambda=0.007$, where $m$ is the number of MCS used
for thermalization at temperature $T(k)$}
\end{figure}

The $f_4$ and $f_5$ functions have the feature of possessing a large number
of minima (for example, $f_4$ has $(2K-1)^N$ local minima in a period). 
The results show again that HSA is a much better alternative when the
number of variables is large, both from the point of view of CPU
time used or the number of function evaluations.
We have chosen the $f_4$ function to 
compare in figure (\ref{fig2}) the evolution of the minimum value 
of the function with the actual number of function evaluations,
for both the FSA and HSA methods, showing again in a different manner
that HSA can find a better
minimum with a less number of function evaluations.
From the results for these functions we infer that in minimization problems 
with a large number of variables and a large number of local minima,
the HSA has the best performance.
Needless to say, we have made our best effort to use the optimal values
for the parameters in each method. It is possible, though, that these values 
could be further improved and the results of tables (\ref{tab1}) and
(\ref{tab2}) slightly modified. We believe, though, that this will
not affect the main conclusions of this paper.

\section{Conclusions}

We have shown by some examples 
how the use of the global update using Hybrid Monte Carlo 
algorithm can indeed improve the performance of simulated annealing methods.
The global updating implicit in HSA allows 
an effective searching scheme and fast annealing schedules and becomes
highly effective, mainly in those problems with a large number of variables
and a large number of metastable minima.

It is clear from the results in the previous section that 
HSA requires in some cases
orders of magnitude less evaluations of the function than 
other methods and can, therefore, give a solutions in less 
computer time. This conclusion remains despite the fact that HSA requires
some extra work when computing the evolution equations since it needs
to compute also the forces $F_i$ acting on the different variables.
In those cases in which the evaluation of the function takes a considerable
amount of computer time, HSA will have an optimal performance, since the
number of function evaluations is greatly reduced as compared to other
simulated annealing methods.
It is conceivable also that 
one could then use efficiently some of the acceleration schemes (Fourier,
wavelet, etc.) available for Monte Carlo methods in order to improve upon
the convergence of the simulated annealing techniques. 
Further developments include 
applying HSA to techniques such as the
Car--Parrinello method for finding the ground state of quantum many
body systems, for which the calculation of the 
energy function is very time consuming. Work on this direction is under
progress.

\noindent ACKNOWLEDGEMENTS

We wish to thank J. Rubio and M. Planas for several discussions.
We acknowledge financial support from DGICyT, grants PB94-1167 and PB94-1172. R.
Salazar is supported by the Agencia Espa\~nola de Cooperaci\'on
Internacional in the Mutis program.

\clearpage

\begin{table}
\begin{center}
\begin{tabular}{|c|c|r|r|r|r|}
\hline
\emph{Function} & \emph{Dimension} &
\emph{FSA} & \emph{VFSR} & \emph{DSA} & \emph{HSA} \\ \hline
$f_1$ & 3 & 480  & 4875 & 79 & 18 \\
$f_1$ & 200 & 8420000 & -- & 474000 & 30 \\
$f_2$ & 2 & 9900 & 1476 & (*) & 165000 \\
$f_3$ & 10 & 2100000 & 319483 & (*) & 720000 \\
$f_4$ & 200 & 12925000  & -- & (*) & 163000 \\
$f_5$ & 10 & 7230000 & -- & 570000 & 118000 \\
\hline
\end{tabular}
\end{center}
\caption{
\label{tab1}
Number of function evaluations averaged over $10$ realizations,
for each of the simulated annealing methods
used for optimization of the different functions to reach the
absolute minimum with an accuracy of $\epsilon=10^{-3}$. 
In those cases marked (*)
it was not possible to reach the absolute minimum. For the HSA, the 
displayed number is the number of function evaluations including 
the calculations of the force.
}
\end{table}

\begin{table}
\begin{center}
\begin{tabular}{|c|c|r|r|r|}
\hline
\emph{Function} & \emph{Dimension} &
\emph{FSA} & \emph{DSA} & \emph{HSA} \\ \hline
$f_1$ & 3 & 0.023 & 0.003 & 0.021 \\
$f_1$ & 200 & 182.898 & 163.763 & 0.039 \\
$f_2$ & 2 & 0.181 & (*) & 5.834 \\
$f_3$ & 10 & 29.454 & (*) & 11.730 \\
$f_4$ & 200 & 1662.177 & (*) & 61.863 \\
$f_5$ & 10 & 119.434 & 13.929 & 4.00 \\
\hline
\end{tabular}
\end{center}

\caption{
\label{tab2}
Similar to table (\ref{tab1}) but showing the 
CPU time (in seconds) needed to reach the absolute minimum with an 
accuracy $\epsilon=10^{-3}$ for each of the simulated annealing
methods explained in the text. All the programs were run on a Silicon Graphics
Origin200 (CPU: R10000 running at 180 MHz, 
Speed: 15.5 SPECfp95). 
}
\end{table}

\clearpage

\clearpage


\begin{thebibliography}{10}
\bibitem{com2}
A similar procedure is used in \cite{ioa96} in the context of sampling Tsallis
  statistics. In their method, the configuration energy difference $\Delta
  E(\x)$ is used instead of the total Hamiltonian difference $\Delta \H(\x)$ as
  the acceptance criterion. In general, this procedure does not properly sample
  the Gibbs distribution, Eq.(\ref{eq:gibbs}), at temperature $T$, although it
  coincides with the one used in this paper as $T \to 0$.
\bibitem{com3}
Time reversibility implies that the original coordinates are exactly recovered
  after numerical integration during a time step if the momenta are reversed.
  Area preserving implies that the Jacobian of the mapping (\ref{eq:leap1}) is
  equal to one.
\end{thebibliography}
\end{document}